\documentclass[9pt,twocolumn,twoside,final]{pnas-new}
\templatetype{pnasresearcharticle}
\title{Transition pathways connecting crystals and quasicrystals}

\author[a]{Jianyuan Yin}
\author[b]{Kai Jiang}
\author[c,1]{An-Chang Shi}
\author[a,1]{Pingwen Zhang}
\author[d,1]{Lei Zhang}
\affil[a]{School of Mathematical Sciences, Laboratory of Mathematics and Applied Mathematics, Peking University, Beijing 100871, China}
\affil[b]{School of Mathematics and Computational Science, Xiangtan University, Hunan 411105, China}
\affil[c]{Department of Physics and Astronomy, McMaster University, Hamilton, Canada L8S 4M1}
\affil[d]{Beijing International Center for Mathematical Research, Center for Quantitative Biology, Peking University, Beijing 100871, China}

\leadauthor{Yin}

\significancestatement{
Despite the fact that tremendous efforts have been made on the study of quasicrystals since their discovery in 1984, nucleation of quasicrystals---the emergence of a quasicrystal from a crystalline phase---still presents an unsolved problem. The difficulties lie in that quasicrystals and crystals are incommensurate structures in general, so there are no obvious epitaxial relations between them. We solved this problem by applying an efficient numerical method to Landau theory of phase transitions and obtained the accurate critical nuclei and transition pathways connecting crystalline and quasicrystalline phases. The proposed computational methodology not only reveals the mechanism of nucleation of quasicrystals, but also paves the way to investigate a wide range of physical problems undergoing the first-order phase transitions.
}

\authorcontributions{Author contributions: P.Z. and L.Z. designed research; J.Y. performed research; K.J. contributed new reagents/analytic tools; J.Y., K.J., A.-C.S., P.Z., and L.Z. analyzed data; and
J.Y., A.-C.S., and L.Z. wrote the paper.}
\authordeclaration{The authors declare no competing interests.}
\correspondingauthor{\textsuperscript{1}To whom correspondence should be addressed. E-mail: zhangl@math.pku.edu.cn or pzhang@pku.edu.cn or shi@mcmaster.ca}


\begin{abstract}
Due to structural incommensurability, the emergence of a quasicrystal from a crystalline phase represents a challenge to computational physics.
Here the nucleation of quasicrystals is investigated by using an efficient computational method applied to a Landau free-energy functional.
Specifically, transition pathways connecting different local minima of the Lifshitz--Petrich model are obtained by using the high-index saddle dynamics.
Saddle points on these paths are identified as the critical nuclei of the 6-fold crystals and 12-fold quasicrystals.
The results reveal that phase transitions between the crystalline and quasicrystalline phases could follow two possible pathways, corresponding to a one-stage phase transition and a two-stage phase transition involving a metastable lamellar quasicrystalline state, respectively.
\end{abstract}

\begin{document}

\maketitle
\thispagestyle{firststyle}
\ifthenelse{\boolean{shortarticle}}{\ifthenelse{\boolean{singlecolumn}}{\abscontentformatted}{\abscontent}}{}

\dropcap{S}ince the discovery of quasicrystals characterized by quasi-periodic positional order with nonclassical rotational symmetries  \cite{shechtman1984metallic}, tremendous progresses have been made on the understanding of these fascinating materials \cite{janot1992quasicrystals, suck2002quasicrystals}.
Various quasicrystals have been reported  \cite{shechtman1984metallic,steurer2004twenty, tsai2008icosahedral, fischer2011colloidal, martinsons2014colloidal}.
Besides examples from metallic alloys, quasicrystalline order has been observed in different systems including Faraday waves and soft matter \cite{edwards1993parametrically, kudrolli1998superlattice, zeng2004supramolecular, hayashida2007polymeric, talapin2009quasicrystalline, zhang2012dodecagonal, huang2018frank, holerca2018dendronized, lindsay2020a15}.
Although the structures of quasicrystals are now well understood \cite{steurer2009crystallography}, the nucleation of quasicrystals, which involves the transition from periodic crystalline structures to quasiperiodic structures, still represents a long-standing unsolved problem.

\begin{figure}[t]
  \centering
  \includegraphics[width=9cm]{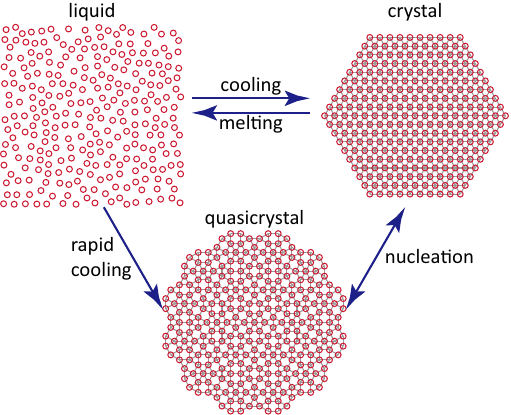}\\
  \caption{Schematic diagram of nucleation and phase transitions between disordered liquid, periodic crystals, and quasicrystals.}
  \label{fig:0}
\end{figure}

In general, nucleation of a stable state from a metastable state could be examined ny using three approaches, i.e. classical nucleation theory, atomistic theory, and density-functional theory \cite{kashchiev2000nucleation, cahn1959free}.
Within the framework of the density-functional theory, the free-energy landscape of the system is described by a free-energy functional determined by the density of molecular species.
Stable and metastable phases of the system correspond to local minima of the free-energy landscape, whereas the minimum energy paths (MEPs) on the free-energy landscape represent the most probable transition pathways between different phases.
Transition states (i.e. index-1 saddle points) on the pathways could be identified as critical nuclei, representing critical states along the transition pathways.
This theoretical framework has been applied successfully to various problems undergoing phase transitions (Fig.~\ref{fig:0})---for instance, the (rapid) cooling of liquids, the melting of a solid, or the nucleation of crystalline structures \cite{zhang2007morphology, lloyd2008localized, cheng2010nucleation, samanta2014microscopic, xu2014nucleation, tang2020atomic}.
However, the study of the phase transition between periodic structures and quasiperiodic structures remains a challenge due to incompatible lattice mismatch.
Thus, a fundamental question in material sciences is: How does a quasicrystalline structure emerge from a crystalline structure?

In this article, we examine the transition pathways connecting quasicrystals and crystals within the framework of density-functional theory.
Specifically, we apply an efficient numerical method based on the high-index saddle dynamics (HiSD) to a Landau free-energy functional, i.e. the Lifshitz--Petrich (LP) model \cite{lifshitz1997theoretical}, with local minima corresponding to two-dimensional (2D) crystalline and quasicrystalline phases.
MEPs connecting various local minima of the model are obtained and critical nuclei of the 6-fold crystalline and 12-fold quasicrystalline states are identified.
In particular, two MEPs connecting two ordered phases are obtained, revealing that the phase transitions between the crystalline and quasicrystalline phases could follow two possible pathways, corresponding to either a one-stage phase transition or a two-stage phase transition involving a metastable intermediate quasicrystalline state, respectively.

\section*{Models and Results}

\paragraph{LP Model}
Although our methodology applies to general free-energy functionals, we will focus on the LP model for simplicity.
The LP model is Landau theory designed to explore quasicrystalline structures with two characteristic wavelength scales \cite{lifshitz1997theoretical}.
Despite its deceivingly simple form, the LP model exhibits a rich phase behavior containing a number of equilibrium ordered phases with 2-, 6-, and 12-fold symmetries \cite{lifshitz1997theoretical, jiang2015stability}.
As such, this simple Landau free-energy provides an ideal model system for the study of transition pathways connecting crystals and quasicrystals.
The LP model assumes a scalar order parameter $\phi(\mathbf{r})$ corresponding to the density profile of the molecules in a volume $V$.
The free-energy functional of the model is given by \cite{lifshitz1997theoretical, jiang2015stability},
\begin{equation}\label{eqn:lp}
\begin{aligned}
  \mathcal{F}(\phi)=\int d\mathbf{r} \;
  \bigg\{\dfrac{1}{2}\left|\left(\nabla^2+1^2\right)\left(\nabla^2+q^2\right)\phi\right|^2 \quad\\
  -\dfrac{\varepsilon}{2}\phi^2-\dfrac{\alpha}{3}\phi^3 + \dfrac{1}{4}\phi^4  \bigg\},
\end{aligned}
\end{equation}
where 1 and $q$ are two characteristic wavelength scales.
The thermodynamic behavior of this model is controlled by two parameters, $\varepsilon$ and $\alpha$,
where $\varepsilon$ is a temperature-like parameter and $\alpha$ is a parameter characterizing the asymmetry of the order parameter \cite{barkan2011stability}.
The coefficients for the spatial derivatives and the $\phi^4$ term could be chosen as $\frac12$ and $\frac14$ by a rescaling of the model (\emph{SI Appendix}) \cite{jiang2015stability}.
Possible equilibrium phases of the model correspond to local minima of the free-energy functional with the mass conservation $\int d\mathbf{r} \;\phi=0$, which are solutions of the Euler--Lagrange equation of the system, $D\mathcal{F}(\phi)=0$.
The Euler--Lagrange equation has multiple solutions, correspond to stable/metastable phases, transition states (critical nuclei, index-1 saddle points), and high-index saddle points of the model system.
In this article, we will focus on the nucleations of two-dimensional 12-fold (dodecagonal) quasicrystals, so a two-dimensional LP model (Eq.~\ref{eqn:lp}) is adopted with $q=2\cos \frac{\pi}{12}$.

The first step of the study is to find accurate stable solutions, corresponding to crystals and quasicrystals, of the Euler--Lagrange equation for the LP free-energy functional Eq.~\ref{eqn:lp}.
Because quasicrystals do not have periodic order, special numerical methods are needed to describe their structures accurately.
In general, discretization methods for quasiperiodic structures include the crystalline approximant method \cite{zhang2008efficient} and the projection method \cite{jiang2014numerical}.
In this article, we adopt the crystalline approximant method to approximate quasiperiodic structures in the whole space with periodic structures in a large domain with proper sizes (\emph{Materials and Methods}).

\begin{figure}[htb]
  \centering
  \includegraphics[width=9cm]{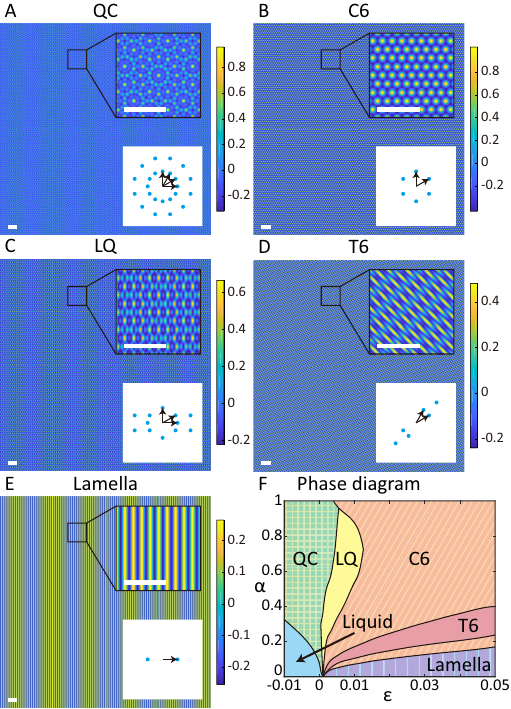}\\
  \caption{
  Stable ordered states and the phase diagram of the 2D LP model with $q=2\cos \frac{\pi}{12}$.
  (\emph{A}) QC at $\varepsilon=5\times10^{-6}$, $\alpha=\sqrt{2}/2$.
  (\emph{B}) C6 at $\varepsilon=0.05$, $\alpha=1$.
  (\emph{C}) LQ at $\varepsilon=0.005$, $\alpha=0.6$.
  (\emph{D}) T6 at $\varepsilon=0.05$, $\alpha=0.3$.
  (\emph{E}) The Lamella state at $\varepsilon=0.05$, $\alpha=0.1$.
  The computational domain is $[0,2\pi L]^2$ with $L=112$, and a $20\pi\times20\pi$ square is zoomed in for better illustration.
  (Scale bars: $10\pi$.)
  Each \emph{Lower Right Inset} is the prominent diffraction pattern in the reciprocal space, and the arrows specify primitive reciprocal vectors (in $\mathbb{Z}$).
  (\emph{F}) Phase diagram of the QC, C6, LQ, T6, Lamella, and Liquid in the $\varepsilon$-$\alpha$ plane.
  }
  \label{fig:1}
\end{figure}

Multiple stable solutions of the Euler--Lagrange equation can be obtained in the LP model with different parameters $\alpha$ and $\varepsilon$.
The initial configurations are composed of plane waves with the appropriate reciprocal wave vectors (\emph{SI Appendix}) \cite{jiang2015stability}.
The simplest solution is $\phi(\mathbf{r})=0$, corresponding to a homogeneous state, i.e. the disordered liquid.
Beside this trivial solution, a number of spatially inhomogeneous solutions, including the 12-fold quasicrystalline state (QC), the 6-fold crystalline state (C6), and the Lamella state have been found for different model parameters.
Interestingly, a lamellar quasicrystalline state (LQ) is identified as a stable state for some parameters as well.
LQ is periodic in one dimension and quasiperiodic in the other dimension, which has been obtained previously in ref. \cite{jiang2020growth} and \cite{subramanian2021exploring}.
Furthermore, a transformed 6-fold crystalline state (T6) that is periodic with less symmetry can also be found in the phase diagram.
The structures of these stable phases are shown in real and reciprocal spaces in Fig.~\ref{fig:1}\emph{A--E}.
It is important to note that the Hessian $D^2 \mathcal{F}(\phi)$ of these states has different multiplicities of zero eigenvalues, corresponding to the numbers of Goldstone modes of these states \cite{goldstone1962broken,shi1999nature} (\emph{Models and Methods}).
The phase diagram of the LP model is constructed and plotted in the $\varepsilon$-$\alpha$ plane in Fig.~\ref{fig:1}\emph{F}, showing stable regions of the QC, C6, LQ, T6, Lamella, and Liquid.
Similar phase diagrams have been obtained by a number of researchers \cite{lifshitz1997theoretical,jiang2015stability,jiang2020growth}.
It is noted that our current study focuses on 2D structures, and thus possible three-dimensional equilibrium phases, such as the body-centered-cubic and gyroid phases, are ignored in the phase diagram Fig.~\ref{fig:1}\emph{F}.

The phase diagram shown in Fig.~\ref{fig:1}\emph{F} is the mean-field phase diagram of the LP model in 2D with $q=2\cos \frac{\pi}{12}$.
Fluctuations could have important effects on the thermodynamics of the model system.
However, we are not aware of any systematic examination of the fluctuation effects on the LP model, and a relevant study might be fluctuation effects on the Landau--Brasovskii model \cite{brazovskii1975phase, swift1977hydrodynamic}.
We would leave the fluctuation effects on the LP model for possible future study.

\paragraph{HiSD Method}
While a stable phase, corresponding to a local minimum of the free-energy functional Eq.~\ref{eqn:lp}, can be calculated by gradient descent algorithms with proper initial configurations, finding a transition state is much more difficult because it does not correspond to a local minimum.
Moreover, multiple zero eigenvalues of its Hessian $D^2 {\cal F}(\phi)$ at stationary points could lead to the degeneracy of transition states.
The problem is further complicated by the fact that there is no a priori knowledge of the transition states.
Most of the existing methods for solving nonlinear equations, such as homotopy methods \cite{chen2004search, mehta2011finding, hao2014bootstrapping} and deflation techniques \cite{brow1971deflation, farrell2015deflation}, are inefficient for this degenerate problem because of superabundant solutions from arbitrary translation.
Surface-walking methods for searching index-1 saddle points, such as the gentlest ascent dynamics \cite{weinan2011gentlest} and dimer-type methods \cite{zhang2016optimization}, are also not capable of computing the transition states from metastable ordered states in the current problem because the eigenvectors with zero eigenvalues of the metastable state would be mistaken as the ascent direction, leading to a failure of escaping the basin of attraction.
On the other hand, the string method \cite{weinan2002string, du2009constrained} with suitable initializations connecting initial and final states could relax to the MEPs.
However, finding initial guesses for a string that is suitable for a particular MEP is not straightforward.
In particular, it is quite difficult in the current problem due to the fact that there are no obvious epitaxial relations between crystals and quasicrystals.
Furthermore, in order to obtain the accurate critical nucleus and MEP, the string method needs a sufficient number of nodes because the critical nucleus is close to the initial metastable state, which could lead to a large increase of the computational cost.
The climbing string method \cite{ren2013climbing} could overcome such difficulty and reduce the computational cost by calculating the half of the MEP from the initial state to the transition state.
However, the climbing string method cannot easily climb out of the basin of attraction because of multiple zero eigenvalues of the initial state.

The presence of zero eigenvalues is computationally a challenge.
To deal with the repeated zero eigenvalues of equilibrium phases in computing the degenerate transition states, we applied a numerical method, HiSD, for high-index saddle points to find degenerate index-1 saddle points, with the inclusion of both negative eigenvalues and zero eigenvalues.
The HiSD for finding index-$k$ saddles ($k$-HiSD) is governed by the following dynamics \cite{yin2019high},
\begin{equation}\label{eqn:hisd}
\dot{\phi}  = -\left(I-2\sum_{j=1}^{k}v_j v_j^\top\right) D\mathcal{F}(\phi),
\end{equation}
where $v_1, \cdots, v_k$ represent the ascent directions, which approximate the eigenvectors corresponding to the smallest $k$ eigenvalues of the Hessian $D^2 \mathcal{F}(\phi)$,
\begin{equation}\label{eqn:hess}
\begin{aligned}
  D^2 \mathcal{F}(\phi)\nu= &\left(\nabla^2+1^2\right)^2\left(\nabla^2+q^2\right)^2\nu \\
  &-\varepsilon\nu +\mathcal{P}(3\phi^2\nu-2\alpha\phi\nu), \quad \forall \nu \;\mathrm{s.t.}  \; \mathcal{P}\nu=\nu.
\end{aligned}
\end{equation}
The LP functional Eq.~\ref{eqn:lp} is highly ill-conditioned because of the eighth-order spatial derivatives, so the locally optimal block preconditioned conjugate gradient (LOBPCG) method \cite{knyazev2001toward} is applied to calculate the smallest $k$ eigenvalues and the corresponding orthonormal eigenvectors.
A preconditioner $\left((\nabla^2+1^2)^2 (\nabla^2+q^2)^2 + \beta I\right)^{-1}$ with $\beta>0$ is applied for better efficiency.

For a metastable state $\phi^*$ whose Hessian $D^2 \mathcal{F}(\phi^*)$ has $m$ zero eigenvalues, we use the LOBPCG method to calculate $\{u^*_1,\cdots,u^*_m\}$ as an orthonormal basis of the nullspace of Hessian $D^2 \mathcal{F}(\phi^*)$ and $u^*_{m+1}$ as a normalized eigenvector of the smallest positive eigenvalue.
Since the smallest positive eigenvalue of Hessians at each stable/metastable ordered state is repeated, there are different choices for $u^*_{m+1}$, which can lead to multiple transition states and MEPs.
Next, we apply the $(m+1)$-HiSD by choosing $\phi(0)=\phi^*+\epsilon u_{m+1}^*$ as the initial search position and $v_i(0)=u^*_i (i=1, \cdots, m+1)$ as the initial ascent directions for searching an index-$(m+1)$ saddle point.
The small positive constant $\epsilon$ is used to push the system away from the minimum, which could be regarded as an upward search on a pathway map \cite{yin2020construction}.
By relaxing the $(m+1)$-HiSD in a semi-implicit scheme for time-dependent $\phi$ with updated ascent directions $v_i$ $(i=1, \cdots, m+1)$ as the eigenvectors at the current position $\phi(t)$ using one-step LOBPCG method, a stationary solution $\phi^{\mathrm{new}}$ can be found, corresponding to a degenerate transition state with only one negative eigenvalue and $m$ repeated zero eigenvalues in most cases.
It should be noted that for non-equilibrium positions $\phi(t)$, the Hessians have no zero eigenvalues in general.
If $\phi^{\mathrm{new}}$ turns out to be a high-index saddle point---for instance, an index-$k$ saddle $(k\leqslant m)$---we then implement $(k-1)$-HiSD to apply a downward search on the pathway map \cite{yin2020construction} to search lower-index saddles.
The initial search position for the downward search is chosen as $\phi(0)=\phi^{\mathrm{new}}+\epsilon u_k^{\mathrm{new}}$, and the initial ascent directions are $v_i=u^{\mathrm{new}}_i (i=1, \cdots, k-1)$, where $\{u^{\mathrm{new}}_1,\cdots,u^{\mathrm{new}}_m\}$ are the orthonormal eigenvectors of $D^2 \mathcal{F}(\phi^{\mathrm{new}})$ calculated by LOBPCG.
This procedure is repeated to new saddle points until the degenerate transition state is located.
The MEP is then obtained by following the gradient flow dynamics along positive and negative unstable directions of the transition state (\emph{SI Appendix}).

\begin{figure}[hbt]
  \centering
  \includegraphics[width=9cm]{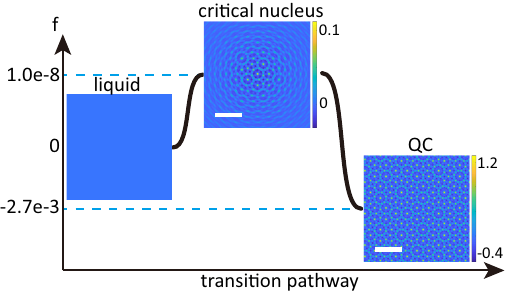}\\
  \caption{Transition pathway from the disordered liquid to QC at $\varepsilon=-0.01$, $\alpha=1$.
  The QC critical nucleus shows a circular shape with a small amplitude.
  (Scale bars: $10\pi$.)}
  \label{fig:2}
\end{figure}

\paragraph{Nucleation from a Liquid to a Quasicrystal}
First, we present the MEP connecting a disordered liquid to a quasicrystal.
By choosing $\varepsilon=-0.01$ and $\alpha=1$, the QC has a lower free-energy density of $f=\mathcal{F}/V=-2.7\times 10^{-3}$ than the disordered liquid with $f=0$.
The critical nucleus of QC from the liquid is shown in Fig.~\ref{fig:2}, which is an index-1 saddle point solution to the Euler--Lagrange equation $D\mathcal{F}(\phi)=0$, corresponding to the transition state on the MEP.
The transition pathway shown in Fig.~\ref{fig:2} represents the possible nucleation process starting from the liquid state toward the quasicrystal.
If the patch of the quasicrystal is smaller than the critical nucleus, it will shrink back to the liquid.
If the patch is larger than the critical nucleus, it will grow and eventually take over the whole system.
The critical nucleus represents a small patch of QC surrounded by damping density waves.
The density wave at the center of the nucleus has a much smaller amplitude than that of the corresponding QC state.
Therefore, the critical nucleus obtained from solving the Euler--Lagrange equation of the system differs significantly from that of the classical nucleation theory.
Along the transition pathway and beyond the critical nucleus, the nucleus grows isotropically with an increasing amplitude at the centre, eventually reaching a full QC phase (see Movie S1 for the liquid $\to$ QC transition pathway).
{Our finding is consistent with the previous results \cite{achim2014growth, jiang2020growth, tang2020atomic}, indicating that there is only one critical nucleus from liquids to quasicrystals, and the growth of quasicrystalline nucleus will fill the whole space. Moreover, our presented example is generic, not limited by special model parameters (see the bifurcation diagram in \emph{SI Appendix}).
}

\begin{figure*}[hbt]
  \centering
  \includegraphics[width=.7\linewidth]{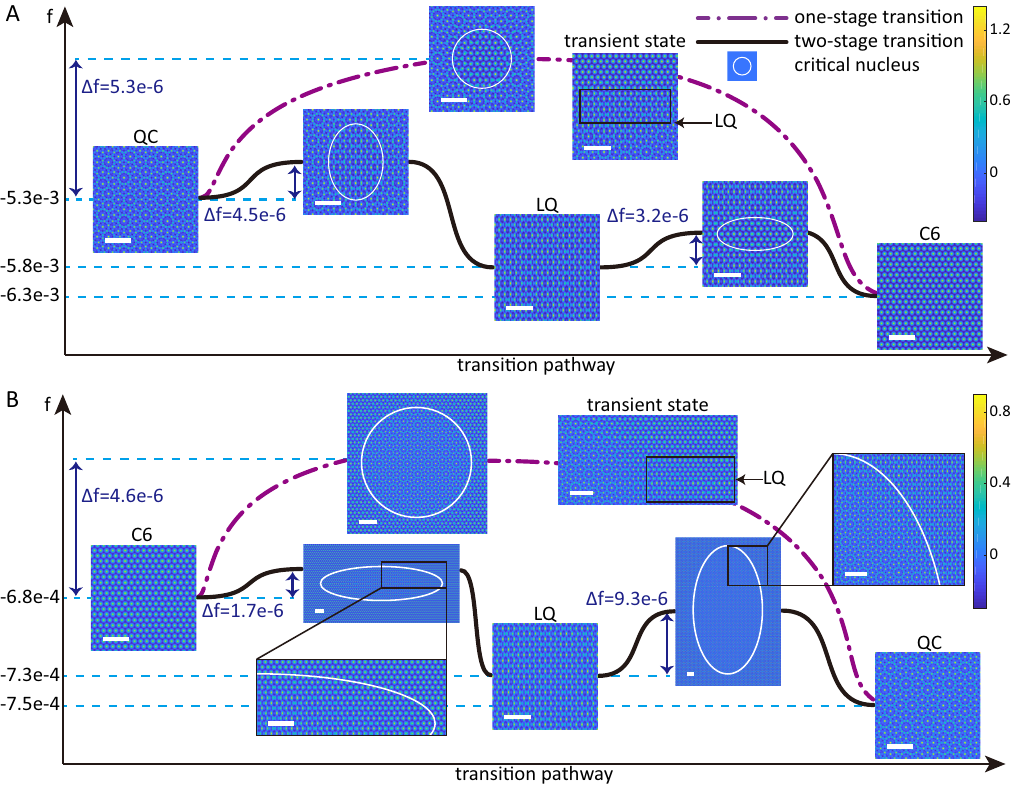}\\
  \caption{Transition pathways between QC and C6.
  (\emph{A}) From QC to C6 at $\varepsilon=0.05$, $\alpha=1$;
  (\emph{B}) From C6 to QC at $\varepsilon=5\times10^{-6}$, $\alpha=\sqrt{2}/2$.
  A one-stage transition pathway (dot-dash curve) includes a circular critical nucleus of QC, and an inserted transient state after nucleation showing LQ serves as an interface between QC and C6.
  A two-stage transition pathway (solid curve) includes a metastable LQ and two ellipsoidal critical nuclei, partially zoomed in for better illustration.
  All critical nuclei are labeled with white rings. (Scale bars: $10\pi$.)
  }
  \label{fig:3}
\end{figure*}

\paragraph{Nucleation from a Quasicrystal to a Crystal}
Next we demonstrate how a quasicrystal would transform to a crystal.
We choose $\varepsilon=0.05$, $\alpha=1$ so that QC is a metastable state with $f=-5.3\times10^{-3}$ and C6 is a stable state with $f=-6.3\times10^{-3}$.
For this case, we found two transition pathways connecting QC to C6.
In the one-stage transition pathway, a circular critical nucleus of C6 shown in Fig.~\ref{fig:3}\emph{A} is observed.
Interestingly, the growing C6 nucleus beyond the critical nucleus shows that a transient state along the transition pathway contains another interphase connecting the C6 and QC states (see Movie S2 for the QC $\to$ C6 transition pathway).
This new interphase is the metastable LQ state, periodic in one dimension and quasiperiodic in the other dimension, which can be stabilized at other parameters (Fig.~\ref{fig:1}\emph{F}).
It is noted that a similar interface between crystals and quasicrystals was obtained with Dirichlet boundary conditions for the phase-field order parameter \cite{cao2021computing}.
This finding indicates that LQ could serve as an intermediate state connecting the QC and C6 phases.
Indeed, a two-stage transition pathway from QC to C6 has been obtained from our calculations. This two-stage pathway reveals a first transition from QC $\to$ LQ and a second transition from LQ $\to$ C6 as shown in Fig.~\ref{fig:3}\emph{A}.
Nucleation at the first stage shows an ellipsoidal critical nucleus of LQ with the periodic direction as the major axis. The energy barrier of the LQ nucleus ($\Delta f=4.5\times10^{-6}$) is lower than the energy barrier of the C6 nucleus ($\Delta f=5.3\times10^{-6}$), indicating that the QC $\to$ LQ transition pathway is the more probable one.
After the QC $\to$ LQ transition, the second-stage transition follows the formation of another ellipsoidal critical nucleus of C6 with the quasiperiodic direction of LQ as the major axis and eventually to the C6 phase (see Movie S3 for the QC $\to$ LQ $\to$ C6 transition pathway).
It is noted that two transition pathways have been observed for the gyroid to lamellar transitions of block copolymers \cite{cheng2010nucleation}.
Furthermore, the appearance of a metastable intermediate state as a precursor of the stable phase is consistent with Ostwald's step rule \cite{ostwald1897studies}.

\paragraph{Nucleation from a Crystal to a Quasicrystal}
Finally, we present results on the emergence of a quasiperiodic structure from a periodic structure.
By choosing $\varepsilon=5\times10^{-6}$ and $\alpha=\sqrt{2}/2$, C6 becomes a metastable state with $f=-6.8\times10^{-4}$ and QC is a stable state with $f=-7.5\times10^{-4}$.
Again, two transition pathways are obtained in this case (Fig.~\ref{fig:3}\emph{B}).
It is noted that, along the one-stage transition pathway, a larger computation domain with $L=306$ was used to avoid the effect of finite domain sizes on growth dynamics.
Similar to the phase transition from QC to C6, a circular critical nucleus of QC is found on the one-stage transition pathway.
After nucleation, the size of the QC nucleus increases with the appearance of the LQ interphase (see Movie S4 for the C6 $\to$ QC transition pathway).
On the other hand, a two-stage transition pathway from C6 to QC via a metastable LQ, as shown in Fig.~\ref{fig:3}\emph{B}, has been obtained.
The critical nucleus of LQ at the first stage assumes an ellipsoidal shape with the quasiperiodic direction as the major axis. The energy barrier of LQ nucleus ($\Delta f=1.7\times10^{-6}$) is lower than that of C6 nucleus ($\Delta f=4.6\times10^{-6}$), indicating that the C6 $\to$ LQ transition would be more likely chosen than the direct C6 $\to$ QC transition pathway.
Because LQ and QC have similar free energies in this case, a small driving force from LQ to QC transition is expected. Therefore, a much larger critical nucleus of QC is found.
A full C6 $\to$ LQ $\to$ QC transition pathway is shown in Movie S5.

\section*{Discussion}
In summary, we applied an efficient numerical method to accurately compute critical nuclei and transition pathways between crystals and quasicrystals.
The computational challenge of the problem stems from the existence of multiple zero eigenvalues of the Hessian of different ordered phases.
We solved this problem by applying the HiSD method to search for high-index saddle points, resulting in degenerate index-1 saddle points corresponding to the critical nuclei.
The proposed methodology is applicable to a wide range of physical problems with degeneracy undergoing phase transitions.
Application of the numerical method to the LP model reveals an interesting set of transition pathways connecting crystalline and quasicrystalline phases.

For the transitions between the crystalline C6 and quasicrystalline QC phases, two transition pathways, corresponding to a one-stage direct transition and a two-stage indirect transition, have been obtained.
We found that a one-dimensional quasicrystalline LQ phase, with periodicity in one direction and quasiperiodicity in the other direction, plays a crucial role to connect C6 with QC.
This discovery is consistent with the phase diagram Fig.~\ref{fig:1}\emph{F}, where the LQ phase can be stabilized between the stable regions of C6 and QC.
The two possible pathways represent different underlying mechanisms of breaking periodicity.
Along the one-stage transition pathway, the periodicity breaks in two dimensions.
On the other hand, for each stage of the two-stage transition pathway, the periodicity is broken along one direction.
Compared with the one-stage transition pathway between C6 and QC, the two-stage transition pathway C6 $\leftrightarrow$ LQ $\leftrightarrow$ QC is, consistent with the Ostwald's step rule, more probable because the LQ nucleus has a lower energy barrier.

Several studies with different models suggested the existence of multiple localized states composed of Hexagon \cite{lloyd2008localized} or quasicrystal \cite{subramanian2018spatially} patches surrounded by the liquid.
This phenomenon generally corresponds to the special case of phase coexistence between the disordered state and the ordered state, which occurs in a narrow region near the phase boundary between these two phases. Thus, the parameter range for the existence of multiple localized states is very limited.
Furthermore, these localized states would correspond to local minima of the free energy landscape, whereas the critical nuclei correspond to saddle points on the free-energy surface.
Using the LP model, we demonstrated that there is only one critical nucleus along the MEP from the liquid state to the quasicrystalline state.
If multiple critical nuclei exist, our method can also find them, and the MEP from the initial state to the final state may pass through multiple transition states, corresponding to multiple energy barriers.
When multiple MEPs exist, the nucleation of quasicrystals would most likely occur along the MEP whose critical state is with the lowest energy barrier.

It should be pointed out that the LP model is a phenomenological model, and the results obtained from such a model cannot be applied directly to physical experiments, unless a connection is made between the physical system and the model parameters.
Nevertheless, these findings shed light on the nucleation and growth of quasicrystals.
The accurate numerical results provide a comprehensive picture of critical nuclei and transition pathways between periodic and quasiperiodic structures.

\matmethods{
\subsection*{Crystalline Approximant Method}
For a given set of $d$ base vectors $\{\mathbf{e}_1^*, \cdots, \mathbf{e}_d^*\}$, a reciprocal lattice vector $\mathbf{k}$ of $d$-dimensional quasicrystals can be expressed as
\begin{equation}\label{eqn:recivec}
  \mathbf{k}=\kappa_1 \mathbf{e}_1^*+\cdots+\kappa_d \mathbf{e}_d^*, \quad \kappa_j \in \mathbb{R}.
\end{equation}
It is important to note that some of the coefficients $\kappa_j$ might be irrational numbers. A quasiperiodic function $\phi(\mathbf{r})$ can be expanded as
\begin{equation}\label{eqn:expand}
  \phi(\mathbf{r})= \sum_{\mathbf{k}}\hat{\phi}(\mathbf{k})\exp (\mathrm{i}\mathbf{k}\cdot\mathbf{r}).
\end{equation}
Since some reciprocal lattice vectors cannot be represented as linear combinations of $\mathbf{e}_i^*$ with integer coefficients, proper rational numbers $L$ are chosen such that $L\kappa_j$ of all the concerned reciprocal lattice vectors $\mathbf{k}$ could be approximated as integers.
As a result, a quasiperiodic function could be approximated by a periodic function with a period $2\pi L$,
\begin{equation}\label{eqn:expand2}
  \phi(\mathbf{r})= \sum_{\mathbf{k}}\hat{\phi}(\mathbf{k})\exp \left(\mathrm{i}\mathbf{k}\cdot\dfrac{\mathbf{r}}{L}\right),
\end{equation}
where $\mathbf{k}$ are linear combinations of $\mathbf{e}_j^*$
with integer coefficients. Within this approximation, the computational domain
becomes $[0,2\pi L]^d$ with periodic boundary conditions for $\phi(\mathbf{r})$.
For the 2D ($d=2$), 12-fold quasicrystals, $q$ is chosen as $2\cos\frac{\pi}{12}$ in the LP model.
For $\mathbf{e}_1^*=(1,0)$ and $\mathbf{e}_2^*=(0,1)$, the coefficients to be approximated are $1, \frac{\sqrt{3}}{2}, \frac{1}{2}, q\cos\frac{\pi}{12}, q\cos\frac{\pi}{4}$, and $q\cos\frac{5\pi}{12}$, and proper values of $L$ are $30$, $82$, $112$, $306$, etc. (\emph{SI Appendix}).
We have tested the accuracy of various $L$ and found that $L\geqslant 112$ gave results within the required accuracy.
Therefore, we set $L=112$ or $306$ in our numerical calculations, and use the spectral methods for Eq.~\ref{eqn:expand2} with $N=1024$ or $3072$ points in each dimension to discretize the order parameter $\phi(\mathbf{r})$.
The stable phases are calculated using gradient flow,
\begin{equation}\label{eqn:var}
  \dot{\phi}=-D\mathcal{F}(\phi)=-\left(\nabla^2+1^2\right)^2\left(\nabla^2+q^2\right)^2\phi
  +\varepsilon\phi-\mathcal{P}(\phi^3-\alpha\phi^2),
\end{equation}
with a semi-implicit scheme \cite{jiang2014numerical}, where $\mathcal{P}\varphi = \varphi - \frac{1}{V}\int d\mathbf{r}\; \varphi$ is the projection operator of the mass conservation constraint.
The nonlinear terms in Eq.~\ref{eqn:var} are treated by using the pseudospectral method \cite{zhang2008efficient}.
The semi-implicit scheme and the pseudospectral method are also applied in the HiSD Eq.~\ref{eqn:hisd}.

\subsection*{Zero Eigenvalues of Hessians at Equilibrium States}
The Hessians of ordered equilibrium states exhibits multiple zero eigenvalues.
We dealt with this by treating the eigenvectors corresponding to repeated zero eigenvalues as unstable directions, and the degenerate transition states (index-1 saddle points) can be calculated from degenerate metastable states using HiSD for index-$k$ saddle points $(k\geqslant1)$.
Here, the (Morse) index of a stationary point of a functional is defined as the number of negative eigenvalues of the Hessian \cite{milnor1963morse}, and the word ``degenerate'' specifies that its Hessian has zero eigenvalues.
For instance, the stable/metastable phases have index 0, and the transition states are index-1 saddle points.
The homogeneous state $\phi(\mathbf r)=0$ is an isolated solution and its Hessian has no zero eigenvalues in general.
For C6, the Hessian has two repeated zero eigenvalues, corresponding to the translation along the $x$ and $y$ axes, while the rotation transformation cannot be realized because of the discretization method.
On the other hand, numerical calculations show that the Hessian at LQ has zero eigenvalues of multiplicity three and the Hessian at QC has zero eigenvalues of multiplicity four.
The various zero-eigenvalue multiplicities of Hessians can be explained with a higher-dimensional description of quasicrystals---that is, a $d$-dimensional quasicrystalline structure can be represented by a projection from a higher-dimensional periodic structure \cite{steurer2009crystallography}.
To calculate QC, a four-dimensional (4D) reciprocal space should be applied in the projection method \cite{jiang2014numerical}, because the reciprocal lattice vectors can be represented by linear combinations of four primitive reciprocal vectors with integer coefficients, as the arrows shown in Fig.~\ref{fig:1}\emph{A--E}.
Since the 2D projection of any 4D translation of QC remains as an equilibrium state, QC is a degenerate solution with four repeated zero eigenvalues.
Correspondingly, three primitive reciprocal vectors are enough to represent LQ with integer coefficients.
In addition, Hessians at nonequilibrium points have no zero eigenvalues generally.
}

\showmatmethods{}

\acknow{This work was supported by National Natural Science Foundation of China Grants 12050002, 21790340, and 11771368; and the Natural Sciences and Engineering Research Council of Canada.
J.Y. was supported by the Elite Program of Computational and Applied Mathematics for Ph.D. Candidates of Peking University.}

\showacknow{}
\bibliography{bib}

\begin{thebibliography}{10}

\bibitem{shechtman1984metallic}
D Shechtman, I Blech, D Gratias, JW Cahn, Metallic phase with long-range
  orientational order and no translational symmetry.
\newblock {\em\protect\JournalTitle{Phys. Rev. Lett.}} \textbf{53}, 1951--1953
  (1984).

\bibitem{janot1992quasicrystals}
C Janot, {\em Quasicrystals: A Primer}.
\newblock (Clarendon Press, Oxford), (1992).

\bibitem{suck2002quasicrystals}
JB Suck, M Schreiber, P H{\"a}ussler, eds., {\em Quasicrystals: An Introduction
  to Structure, Physical Properties and Applications}.
\newblock (Springer, Berlin), (2002).

\bibitem{steurer2004twenty}
W Steurer, Twenty years of structure research on quasicrystals. {P}art {I}.
  {P}entagonal, octagonal, decagonal and dodecagonal quasicrystals.
\newblock {\em\protect\JournalTitle{Z. Kristallogr.}} \textbf{219}, 391--446
  (2004).

\bibitem{tsai2008icosahedral}
AP Tsai, Icosahedral clusters, icosaheral order and stability of
  quasicrystals{\textemdash}a view of metallurgy.
\newblock {\em\protect\JournalTitle{Sci. Technol. Adv. Mater.}} \textbf{9},
  013008 (2008).

\bibitem{fischer2011colloidal}
S Fischer, et~al., Colloidal quasicrystals with 12-fold and 18-fold diffraction
  symmetry.
\newblock {\em\protect\JournalTitle{Proc. Natl. Acad. Sci. U.S.A.}}
  \textbf{108}, 1810--1814 (2011).

\bibitem{martinsons2014colloidal}
M Martinsons, M Sandbrink, M Schmiedeberg, Colloidal trajectories in
  two-dimensional light-induced quasicrystals with 14-fold symmetry due to
  phasonic drifts.
\newblock {\em\protect\JournalTitle{Acta Phys. Pol. A}} \textbf{126}, 568--571
  (2014).

\bibitem{edwards1993parametrically}
WS Edwards, S Fauve, Parametrically excited quasicrystalline surface waves.
\newblock {\em\protect\JournalTitle{Phys. Rev. E}} \textbf{47}, R788--R791
  (1993).

\bibitem{kudrolli1998superlattice}
A Kudrolli, B Pier, JP Gollub, Superlattice patterns in surface waves.
\newblock {\em\protect\JournalTitle{Physica D}} \textbf{123}, 99--111 (1998).

\bibitem{zeng2004supramolecular}
X Zeng, et~al., Supramolecular dendritic liquid quasicrystals.
\newblock {\em\protect\JournalTitle{Nature}} \textbf{428}, 157--160 (2004).

\bibitem{hayashida2007polymeric}
K Hayashida, T Dotera, A Takano, Y Matsushita, Polymeric quasicrystal:
  Mesoscopic quasicrystalline tiling in {$ABC$} star polymers.
\newblock {\em\protect\JournalTitle{Phys. Rev. Lett.}} \textbf{98}, 195502
  (2007).

\bibitem{talapin2009quasicrystalline}
DV Talapin, et~al., Quasicrystalline order in self-assembled binary
  nanoparticle superlattices.
\newblock {\em\protect\JournalTitle{Nature}} \textbf{461}, 964--967 (2009).

\bibitem{zhang2012dodecagonal}
J Zhang, FS Bates, Dodecagonal quasicrystalline morphology in a
  poly(styrene-b-isoprene-b-styrene-b-ethylene oxide) tetrablock terpolymer.
\newblock {\em\protect\JournalTitle{J. Am. Chem. Soc.}} \textbf{134},
  7636--7639 (2012).

\bibitem{huang2018frank}
M Huang, et~al., Frank--{K}asper and related quasicrystal spherical phases in
  macromolecules.
\newblock {\em\protect\JournalTitle{Sci. China Chem.}} \textbf{61}, 33--45
  (2018).

\bibitem{holerca2018dendronized}
MN Holerca, et~al., Dendronized poly(2-oxazoline) displays within only five
  monomer repeat units liquid quasicrystal, a15 and $\sigma$ {F}rank--{K}asper
  phases.
\newblock {\em\protect\JournalTitle{J. Am. Chem. Soc.}} \textbf{140},
  16941--16947 (2018).

\bibitem{lindsay2020a15}
AP Lindsay, et~al., A15, $\sigma$, and a quasicrystal: Access to complex
  particle packings via bidisperse diblock copolymer blends.
\newblock {\em\protect\JournalTitle{ACS Macro Lett.}} \textbf{9}, 197--203
  (2020).

\bibitem{steurer2009crystallography}
W Steurer, S Deloudi, {\em Crystallography of Quasicrystals: Concepts, Methods
  and Structures}.
\newblock (Springer, Berlin), (2009).

\bibitem{kashchiev2000nucleation}
D Kashchiev, {\em Nucleation: Basic Theory with Applications}.
\newblock (Butterworth-Heinemann, Oxford), (2000).

\bibitem{cahn1959free}
JW Cahn, JE Hilliard, Free energy of a nonuniform system. iii. nucleation in a
  two-component incompressible fluid.
\newblock {\em\protect\JournalTitle{J. Chem. Phys.}} \textbf{31}, 688--699
  (1959).

\bibitem{zhang2007morphology}
L Zhang, LQ Chen, Q Du, Morphology of critical nuclei in solid-state phase
  transformations.
\newblock {\em\protect\JournalTitle{Phys. Rev. Lett.}} \textbf{98}, 265703
  (2007).

\bibitem{lloyd2008localized}
DJB Lloyd, B Sandstede, D Avitabile, AR Champneys, Localized hexagon patterns
  of the planar {S}wift--{H}ohenberg equation.
\newblock {\em\protect\JournalTitle{SIAM J. Appl. Dyn. Syst.}} \textbf{7},
  1049--1100 (2008).

\bibitem{cheng2010nucleation}
X Cheng, L Lin, W E, P Zhang, AC Shi, Nucleation of ordered phases in block
  copolymers.
\newblock {\em\protect\JournalTitle{Phys. Rev. Lett.}} \textbf{104}, 148301
  (2010).

\bibitem{samanta2014microscopic}
A Samanta, ME Tuckerman, TQ Yu, W E, Microscopic mechanisms of equilibrium
  melting of a solid.
\newblock {\em\protect\JournalTitle{Science}} \textbf{346}, 729--732 (2014).

\bibitem{xu2014nucleation}
X Xu, CL Ting, I Kusaka, ZG Wang, Nucleation in polymers and soft matter.
\newblock {\em\protect\JournalTitle{Annu. Rev. Phys. Chem.}} \textbf{65},
  449--475 (2014).

\bibitem{tang2020atomic}
S Tang, et~al., An atomic scale study of two-dimensional quasicrystal
  nucleation controlled by multiple length scale interactions.
\newblock {\em\protect\JournalTitle{Soft Matter}} \textbf{16}, 5718--5726
  (2020).

\bibitem{lifshitz1997theoretical}
R Lifshitz, DM Petrich, Theoretical model for {F}araday waves with
  multiple-frequency forcing.
\newblock {\em\protect\JournalTitle{Phys. Rev. Lett.}} \textbf{79}, 1261--1264
  (1997).

\bibitem{jiang2015stability}
K Jiang, J Tong, P Zhang, AC Shi, Stability of two-dimensional soft
  quasicrystals in systems with two length scales.
\newblock {\em\protect\JournalTitle{Phys. Rev. E}} \textbf{92}, 042159 (2015).

\bibitem{barkan2011stability}
K Barkan, H Diamant, R Lifshitz, Stability of quasicrystals composed of soft
  isotropic particles.
\newblock {\em\protect\JournalTitle{Phys. Rev. B}} \textbf{83}, 172201 (2011).

\bibitem{zhang2008efficient}
P Zhang, X Zhang, An efficient numerical method of {L}andau--{B}razovskii
  model.
\newblock {\em\protect\JournalTitle{J. Comput. Phys.}} \textbf{227}, 5859--5870
  (2008).

\bibitem{jiang2014numerical}
K Jiang, P Zhang, Numerical methods for quasicrystals.
\newblock {\em\protect\JournalTitle{J. Comput. Phys.}} \textbf{256}, 428--440
  (2014).

\bibitem{jiang2020growth}
Z Jiang, S Quan, N Xu, L He, Y Ni, Growth modes of quasicrystals involving
  intermediate phases and a multistep behavior studied by phase field crystal
  model.
\newblock {\em\protect\JournalTitle{Phys. Rev. Materials}} \textbf{4}, 023403
  (2020).

\bibitem{subramanian2021exploring}
P Subramanian, IG Kevrekidis, PG Kevrekidis, Exploring critical points of
  energy landscapes: From low-dimensional examples to phase field crystal
  {PDEs}.
\newblock {\em\protect\JournalTitle{Commun. Nonlinear Sci. Numer. Simulat.}}
  \textbf{96}, 105679 (2021).

\bibitem{goldstone1962broken}
J Goldstone, A Salam, S Weinberg, Broken symmetries.
\newblock {\em\protect\JournalTitle{Phys. Rev.}} \textbf{127}, 965--970 (1962).

\bibitem{shi1999nature}
AC Shi, Nature of anisotropic fluctuation modes in ordered systems.
\newblock {\em\protect\JournalTitle{J. Phys. Condens. Matter}} \textbf{11},
  10183--10197 (1999).

\bibitem{brazovskii1975phase}
SA Brazovskii, Phase transition of an isotropic system to a nonuniform state.
\newblock {\em\protect\JournalTitle{Zh. Eksp. Teor. Fiz.}} \textbf{68},
  175--185 (1975).

\bibitem{swift1977hydrodynamic}
J Swift, PC Hohenberg, Hydrodynamic fluctuations at the convective instability.
\newblock {\em\protect\JournalTitle{Phys. Rev. A}} \textbf{15}, 319--328
  (1977).

\bibitem{chen2004search}
C Chen, Z Xie, Search extension method for multiple solutions of a nonlinear
  problem.
\newblock {\em\protect\JournalTitle{Comput. Math. with Appl.}} \textbf{47},
  327--343 (2004).

\bibitem{mehta2011finding}
D Mehta, Finding all the stationary points of a potential-energy landscape via
  numerical polynomial-homotopy-continuation method.
\newblock {\em\protect\JournalTitle{Phys. Rev. E}} \textbf{84}, 025702 (2011).

\bibitem{hao2014bootstrapping}
W Hao, JD Hauenstein, B Hu, AJ Sommese, A bootstrapping approach for computing
  multiple solutions of differential equations.
\newblock {\em\protect\JournalTitle{J. Comput. Appl. Math.}} \textbf{258},
  181--190 (2014).

\bibitem{brow1971deflation}
KM Brow, WB Gearhart, Deflation techniques for the calculation of further
  solutions of a nonlinear system.
\newblock {\em\protect\JournalTitle{Numer. Math.}} \textbf{16}, 334--342
  (1971).

\bibitem{farrell2015deflation}
PE Farrell, {\'{A}} Birkisson, SW Funke, Deflation techniques for finding
  distinct solutions of nonlinear partial differential equations.
\newblock {\em\protect\JournalTitle{SIAM J. Sci. Comput.}} \textbf{37},
  A2026--A2045 (2015).

\bibitem{weinan2011gentlest}
W E, X Zhou, The gentlest ascent dynamics.
\newblock {\em\protect\JournalTitle{Nonlinearity}} \textbf{24}, 1831--1842
  (2011).

\bibitem{zhang2016optimization}
L Zhang, Q Du, Z Zheng, Optimization-based shrinking dimer method for finding
  transition states.
\newblock {\em\protect\JournalTitle{SIAM J. Sci. Comput.}} \textbf{38},
  A528--A544 (2016).

\bibitem{weinan2002string}
W E, W Ren, E Vanden-Eijnden, String method for the study of rare events.
\newblock {\em\protect\JournalTitle{Phys. Rev. B}} \textbf{66}, 052301 (2002).

\bibitem{du2009constrained}
Q Du, L Zhang, A constrained string method and its numerical analysis.
\newblock {\em\protect\JournalTitle{Commun. Math. Sci.}} \textbf{7}, 1039--1051
  (2009).

\bibitem{ren2013climbing}
W Ren, E Vanden-Eijnden, A climbing string method for saddle point search.
\newblock {\em\protect\JournalTitle{J. Chem. Phys.}} \textbf{138}, 134105
  (2013).

\bibitem{yin2019high}
J Yin, L Zhang, P Zhang, High-index optimization-based shrinking dimer method
  for finding high-index saddle points.
\newblock {\em\protect\JournalTitle{SIAM J. Sci. Comput.}} \textbf{41},
  A3576--A3595 (2019).

\bibitem{knyazev2001toward}
AV Knyazev, Toward the optimal preconditioned eigensolver: Locally optimal
  block preconditioned conjugate gradient method.
\newblock {\em\protect\JournalTitle{SIAM J. Sci. Comput.}} \textbf{23},
  517--541 (2001).

\bibitem{yin2020construction}
J Yin, Y Wang, JZY Chen, P Zhang, L Zhang, Construction of a pathway map on a
  complicated energy landscape.
\newblock {\em\protect\JournalTitle{Phys. Rev. Lett.}} \textbf{124}, 090601
  (2020).

\bibitem{achim2014growth}
CV Achim, M Schmiedeberg, H L{\"o}wen, Growth modes of quasicrystals.
\newblock {\em\protect\JournalTitle{Phys. Rev. Lett.}} \textbf{112}, 255501
  (2014).

\bibitem{cao2021computing}
D Cao, J Shen, J Xu, Computing interface with quasiperiodicity.
\newblock {\em\protect\JournalTitle{J. Comput. Phys.}} \textbf{424}, 109863
  (2021).

\bibitem{ostwald1897studies}
W Ostwald, Studies on the formation and change of solid matter.
\newblock {\em\protect\JournalTitle{Z. Phys. Chem.}} \textbf{22}, 289--330
  (1897).

\bibitem{subramanian2018spatially}
P Subramanian, AJ Archer, E Knobloch, AM Rucklidge, Spatially localized
  quasicrystalline structures.
\newblock {\em\protect\JournalTitle{New J. Phys.}} \textbf{20}, 122002 (2018).

\bibitem{milnor1963morse}
J Milnor, {\em Morse Theory}.
\newblock (Princeton University Press, Princeton, NJ), (1963).

\end{thebibliography}


\begin{thebibliography}{1}

\bibitem{lifshitz1997theoretical}
R Lifshitz, DM Petrich, Theoretical model for {F}araday waves with
  multiple-frequency forcing.
\newblock {\em\protect\JournalTitle{Phys. Rev. Lett.}} \textbf{79}, 1261--1264
  (1997).

\bibitem{jiang2015stability}
K Jiang, J Tong, P Zhang, AC Shi, Stability of two-dimensional soft
  quasicrystals in systems with two length scales.
\newblock {\em\protect\JournalTitle{Phys. Rev. E}} \textbf{92}, 042159 (2015).

\bibitem{subramanian2016three}
P Subramanian, AJ Archer, E Knobloch, AM Rucklidge, Three-dimensional
  icosahedral phase field quasicrystal.
\newblock {\em\protect\JournalTitle{Phys. Rev. Lett.}} \textbf{117}, 075501
  (2016).

\bibitem{subramanian2018spatially}
P Subramanian, AJ Archer, E Knobloch, AM Rucklidge, Spatially localized
  quasicrystalline structures.
\newblock {\em\protect\JournalTitle{New J. Phys.}} \textbf{20}, 122002 (2018).

\bibitem{subramanian2021exploring}
P Subramanian, IG Kevrekidis, PG Kevrekidis, Exploring critical points of
  energy landscapes: From low-dimensional examples to phase field crystal
  {PDEs}.
\newblock {\em\protect\JournalTitle{Commun. Nonlinear Sci. Numer. Simulat.}}
  \textbf{96}, 105679 (2021).

\bibitem{barkan2011stability}
K Barkan, H Diamant, R Lifshitz, Stability of quasicrystals composed of soft
  isotropic particles.
\newblock {\em\protect\JournalTitle{Phys. Rev. B}} \textbf{83}, 172201 (2011).

\bibitem{davenport1946simultaneous}
H Davenport, K Mahler, Simultaneous {D}iophantine approximation.
\newblock {\em\protect\JournalTitle{Duke Math. J.}} \textbf{13}, 105--111
  (1946).

\bibitem{jiang2014numerical}
K Jiang, P Zhang, Numerical methods for quasicrystals.
\newblock {\em\protect\JournalTitle{J. Comput. Phys.}} \textbf{256}, 428--440
  (2014).

\bibitem{weinan2002string}
W E, W Ren, E Vanden-Eijnden, String method for the study of rare events.
\newblock {\em\protect\JournalTitle{Phys. Rev. B}} \textbf{66}, 052301 (2002).

\end{thebibliography}

\end{document}


\maketitle
\SItext

\subsection*{Dimensionless form of the two-dimensional Lifshitz--Petrich model}
The original two-dimensional Lifshitz--Petrich (LP) model may contain seven parameters $c>0, q_2\geqslant q_1>0, \beta>0$ and $\varepsilon, \alpha, A \in\mathbb{R}$ for a general physical system \cite{lifshitz1997theoretical},
\begin{equation}\label{eqn:lp1}
\begin{aligned}
  \mathcal{F}(\phi)=&\int d\mathbf{r} \;
  \bigg\{\dfrac{c}{2}\left|\left(\nabla^2+q_1^2\right)\left(\nabla^2+q_2^2\right)\phi\right|^2
  -\dfrac{\varepsilon}{2}\phi^2-\dfrac{\alpha}{3}\phi^3 + \dfrac{\beta}{4}\phi^4  \bigg\},\\
  \mathrm{s.t.}\;&\int d\mathbf{r} \;(\phi-A)=0,
\end{aligned}
\end{equation}
but only three parameters remains after rescaling \cite{jiang2015stability}.
Let $q=q_2/q_1$ and $\phi(\mathbf{r}) = q_1^4\sqrt{\beta^{-1}c}\tilde{\phi}(q_1 \mathbf{r})+A$, we have
\begin{equation}\label{eqn:lp2}
\begin{aligned}
  \tilde{\mathcal{F}}(\tilde{\phi})=\beta c^{-2} q_1^{-14}\left(\mathcal{F}(\phi)-\tilde{C}\right)=
  &\int d\mathbf{r} \;
  \bigg\{\dfrac{1}{2}\left|\left(\nabla^2+ 1^2\right)\left(\nabla^2+q^2\right)\tilde{\phi}\right|^2
  -\dfrac{\tilde{\varepsilon}}{2}\tilde{\phi}^2-\dfrac{\tilde{\alpha}}{3}\tilde{\phi}^3 + \dfrac{1}{4}\tilde{\phi}^4 \bigg\},\\
  \mathrm{s.t.}\;&\int d\mathbf{r} \;\tilde{\phi}=0,
\end{aligned}
\end{equation}
where the rescaled parameters are ($V$ is the integral area in Eq.~\ref{eqn:lp1})
\begin{equation}\label{eqn:lp3}
  \tilde{\alpha}=\dfrac{\alpha-3A\beta}{\sqrt{\beta c}q_1^4}, \quad
  \tilde{\varepsilon}=\dfrac{\varepsilon+2A\alpha-3A^2\beta}{cq_1^8}, \quad
  \tilde{C}=V\left( \dfrac{c}{2}q_1^4q_2^4A^2- \dfrac{\varepsilon}{2}A^2- \dfrac{\alpha}{3}A^3+\dfrac{\beta}{4}A^4\right).
\end{equation}
Therefore, without loss of generality, we can set $c=\beta=q_1=1$ and $A=0$ in the LP model.

The parameter $q$ represents the ratio of two characteristic wavelength scales in the system.
To stabilize different quasicrystals, the LP model requires certain choices of $q$.
In the current study, $q$ is chosen as $2\cos \frac{\pi}{12}$ to stabilize various ordered phases of interest, including the two-dimensional 12-fold (dodecagonal) quasicrystals.
For the three-dimensional icosahedral and two-dimensional 10-fold (decagonal) quasicrystals, $q$ is chosen as $2\cos \frac{\pi}{5}$.
For the two-dimensional 8-fold (octagonal) quasicrystals, $q$ is chosen as $2\cos \frac{\pi}{8}$ \cite{jiang2015stability}.
A similar phase field crystal model was used in \cite{subramanian2016three,subramanian2018spatially,subramanian2021exploring} to stabilize quasicrystals with various symmetries, which is slightly different from Eq.~\ref{eqn:lp1}, and our methodology can be directly applied to this model as well.

The actual free energy barrier is proportional to the values $\Delta\mathcal{F}$ obtained in our calculations that is in unit of $kT$, that is, $\Delta F_{\mathrm{physical}} = \Delta \mathcal{F} \;\xi \;kT$, where $\xi$ is a model-dependent parameter.
From the derivation of the Landau theory from certain molecular models, all the parameters in the Landau theory were specified by the molecular parameters \cite{barkan2011stability}.
Since we have taken the Landau theory as a generic model of phase transitions, the information about the exact value of $\xi$ is not available within our current approach.
If we take the Landau theory specified by Eq.~\ref{eqn:lp1} as the starting physical model, we could determine the value of $\xi$ via the scaling analysis following Eq.~\ref{eqn:lp1}.
Specifically, the free energy given by Eq.~\ref{eqn:lp1} is in the unit of $kT$.
Thus, the scaled free energy is in the unit of $\beta^{-1}c^2q_1^{14}kT$, or $\xi = \beta^{-1}c^2q_1^{14}$.

\subsection*{Proper values of computational domain sizes}
\begin{figure}
\centering
\includegraphics[width=\textwidth]{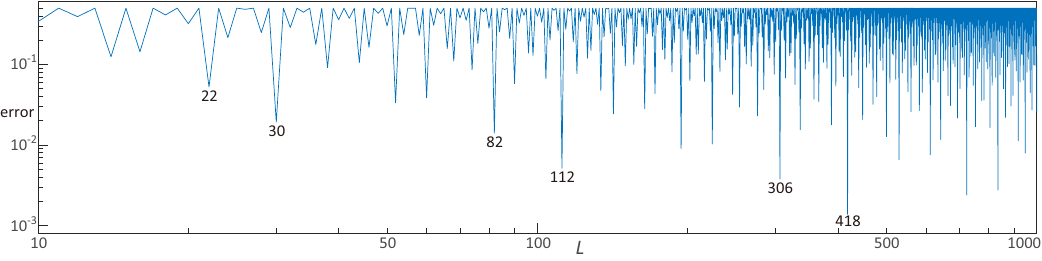}
\caption{The approximate errors of different computational domain sizes $[0,2\pi L]^2$ for the two-dimensional 12-fold quasicrystals.
Both the horizontal ($L$) and vertical (approximate error Eq.~\ref{eqn:sdaae}) axes are shown with a logarithmic scale.
Proper values of $L$ are 22, 30, 82, 112, 306, 418, etc.
}
\label{fig:s2sda}
\end{figure}

For the two-dimensional 12-fold quasicrystals, the reciprocal vectors should include,
\begin{equation}\label{eqn:sdarv}
\begin{aligned}
  \mathbf{k}_j&=\left(\cos \dfrac{j\pi }{6}, \sin \dfrac{j\pi}{6}\right), \quad &&j=1,\cdots,12,\\
  \mathbf{k}_{j}&=\left(q\cos \dfrac{(2j-1)\pi }{12}, q\sin \dfrac{(2j-1)\pi}{12}\right), \quad &&j=13,\cdots,24,
\end{aligned}
\end{equation}
where $q= 2 \cos \frac{\pi}{12}$.
For $\mathbf{e}_1^*=(1,0)$ and $\mathbf{e}_2^*=(0,1)$ in the crystalline approximant method, the coefficients to be approximated simultaneously are $1, \frac{\sqrt{3}}{2}, \frac{1}{2}, q\cos\frac{\pi}{12}, q\cos\frac{\pi}{4}, q\cos\frac{5\pi}{12}$.
The approximate error of the computational domain $[0,2\pi L]^2$ caused by simultaneous Diophantine approximation \cite{davenport1946simultaneous} is,
\begin{equation}\label{eqn:sdaae}
\left\| L\left(1, \frac{\sqrt{3}}{2}, \frac{1}{2}, q\cos\frac{\pi}{12}, q\cos\frac{\pi}{4}, q\cos\frac{5\pi}{12}\right) -
\left(\left[L\right], \left[\frac{\sqrt{3}L}{2}\right], \left[\frac{L}{2}\right], \left[qL\cos\frac{\pi}{12}\right], \left[qL\cos\frac{\pi}{4}\right], \left[qL\cos\frac{5\pi}{12}\right]\right)\right\|_\infty,
\end{equation}
where $[a]$ rounds $a$ to the nearest integer.
The approximate errors Eq.~\ref{eqn:sdaae} of integers $10\leqslant L\leqslant 1000$ are shown in Fig.~\ref{fig:s2sda}.
The proper values of $L$, with a smaller approximate error Eq.~\ref{eqn:sdaae} than all smaller domains, can be obtained directly from Fig.~\ref{fig:s2sda} as $L=22, 30, 82, 112, 306, 418$ etc., with approximate errors shown in Table~\ref{tab:sda}.

\subsection*{Initial values for the (meta)stable states}
The free energy functional of the LP model possesses many local minima, corresponding to the stable and metastable states of the system.
The stable state (C6, LQ, QC, etc.) of the system is the state which has the lowest free energy at a given set of $\varepsilon$ and $\alpha$.
When these parameters are changed, the stable state may become another equilibrium state.
The overall phase behaviour of the system is shown in the phase diagram (Fig.~2F), which displays the stability region of each structure in the phase space spanned by $\varepsilon$ and $\alpha$.

Locating the local minima of the model free energy functional depends on the initial configuration.
In order to find a particular state, we used initial configurations composed of plane waves with the correct reciprocal wave vectors.
For the computational domain size $[0,2\pi L]^2$ in the crystalline approximant method, the reciprocal vector $\mathbf{k}_j$ is approximated by $\mathbf{k}_{j,L}= L^{-1}[L\mathbf{k}_{j}]$, where $[\mathbf{k}]$ rounds each entry of $\mathbf{k}$ to the nearest integer.
The initial values of these phases are prepared as follows \cite{jiang2014numerical},
\begin{equation}\label{eqn:initialvalue}
\phi(\mathbf{r}) = \sum_{j\in J} a\exp \left(\mathrm{i}\mathbf{k}_{j,L}\cdot \mathbf{r}\right).
\end{equation}
In practical computation, $a$ is taken as 0.058 for better convergent performance.
The initial choice of the set $J$, which is based on the symmetry of the intended structure, is taken as follows:
\begin{enumerate}
  \item QC: $J=\{1,2,3,\cdots,24\}$;
  \item C6: $J=\{1,3,5,\cdots,11\}$;
  \item LQ: $J=\{1,3,5,6,7,9,11,12,13,18,19,24\}$;
  \item T6: $J=\{1,2,7,8,14,20\}$;
  \item Lamella: $J=\{1,2\}$.
\end{enumerate}
Then the (meta)stable states can be obtained using gradient flow (Eq.~7 in the main text) with a semi-implicit scheme.
Note that the (meta)stable states have much more reciprocal vectors with nonzero Fourier coefficients than the initial state, while the amplitudes of the Fourier coefficients are smaller.

\subsection*{Computing the minimum energy path}
The minimum energy pathways (MEPs) are the most probable pathways of the corresponding overdamped Langevin equation from the initial metastable state to the final stable state \cite{weinan2002string}, as shown in Fig.~3--4 and Movies S1--5.
The MEP consists of the nucleation part (energy-increasing dynamics from the initial state to the transition state) and the growth (nucleus growing) part (energy-decreasing dynamics from the transition state to the final state).
Both the nucleation part and the growth part correspond to gradient flows from the transition state to a minimum.
Therefore, the MEP between minima can be obtained directly using gradient flows from the corresponding transition state.

Assuming that the transition state $\phi_0$ is obtained, we first calculate the unstable direction $u_0$ as the eigenvector corresponding to the negative eigenvalue of the Hessian $D^2 \mathcal{F}(\phi_0)$.
We solve gradient flow dynamics from $\phi_0+\epsilon u_0$ and $\phi_0-\epsilon u_0$ until convergence to the initial and final states, where $0<\epsilon \ll 1$.
Then the first part of the MEP is the time-reverse gradient flow dynamics from the transition state to the initial state.
The second part of the MEP is the gradient flow dynamics from the transition state to the final state.
For the two-stage transition pathways in Fig.~4, each stage is calculated as above, and the entire MEP is the combination of two stages.

\subsection*{Bifurcation diagram of the transition from the liquid to the quasicrystal}

\begin{figure}
\centering
\includegraphics[width=\textwidth]{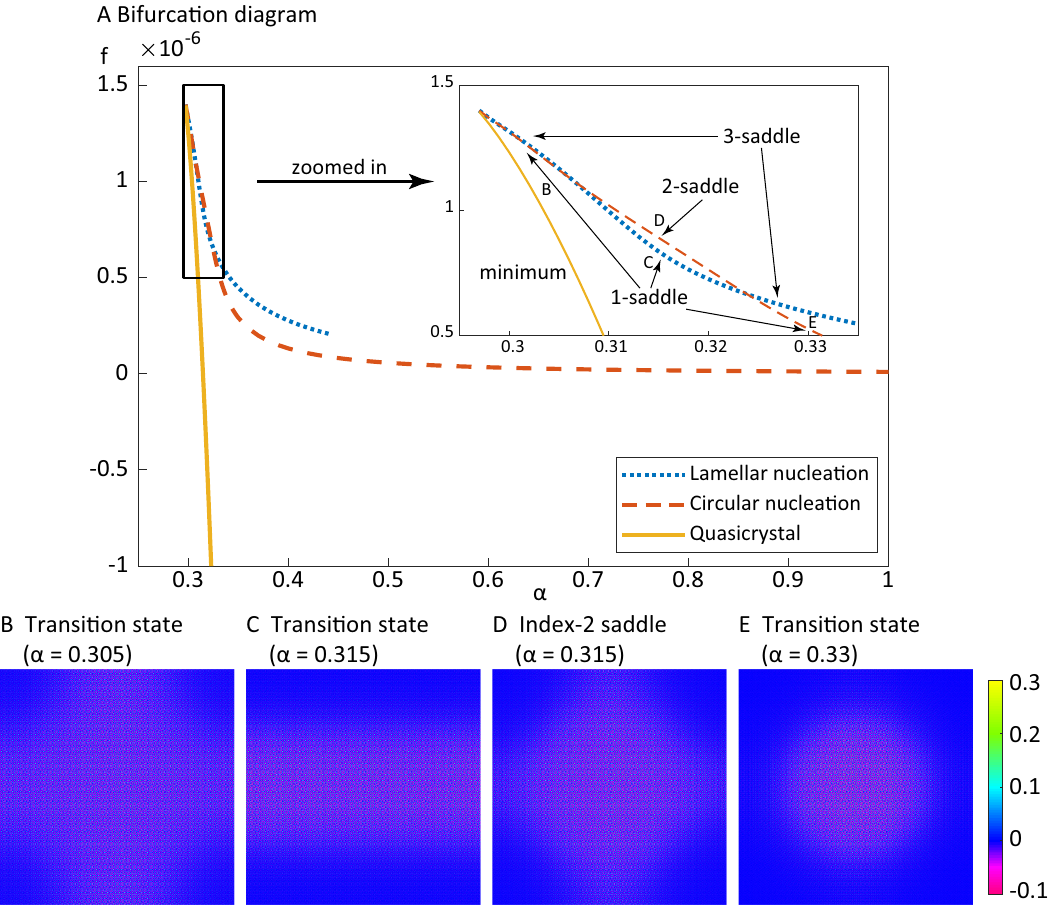}
\caption{
{
The transition state from the liquid state to the quasicrystal state in the LP model with mass conservation with $\varepsilon=-0.01$.
  \emph{A} The bifurcation diagram of the transition state. As $\alpha$ increases, the transition state between the liquid state and the quasicrystal state changes from a circular nucleation state to a lamellar nucleation state, and to a circular nucleation state.
  Yellow solid line: Quasicrystal metastable state.
  Red dashed line: Circular nucleation state.
  Blue dotted line: Lamellar nucleation state.
  \emph{B} The transition state (index-1 saddle point) at $\alpha=0.305$ is a circular nucleation state, and the quasicrystalline pattern is the dominant structure of this saddle point.
  \emph{C} The transition state at $\alpha=0.315$ is a lamellar nucleation state.
  \emph{D} The circular nucleation state becomes an index-2 saddle point at $\alpha=0.315$.
  \emph{E} The transition state at $\alpha=0.33$ is a circular nucleation state, and the liquid state is the dominant structure of this saddle point.
}}
\label{fig:s1}
\end{figure}

{For the nucleation from a liquid state to a quasicrystal state, we present the result at $\varepsilon=-0.01$, $\alpha = 1$ in the main text.
To further demonstrate the generality of our results, we presented the bifurcation diagram of the transition state here.
We fix $\varepsilon=-0.01$ and increase $\alpha$ from $0$ to $1$, with the domain size $L=112$ and the mesh number $N=1024$.
The quasicrystalline state (the yellow solid line in Fig.~\ref{fig:s1}A) emerges as a local minimum at around $\alpha \approx0.297$.
Meanwhile, a circular nucleation state (the red dashed line in Fig.~\ref{fig:s1}A) and a lamellar nucleation state (the blue dotted line in Fig.~\ref{fig:s1}A) also emerge.
At the beginning of the emergence of the quasicrystal metastable state, the circular nucleation state is the transition state (Fig.~\ref{fig:s1}B) between the liquid state and the quasicrystal state, where the quasicrystalline pattern is the dominant structure.
After a transcritical bifurcation, the lamellar nucleation state becomes the transition state (Fig.~\ref{fig:s1}C) between the liquid state and the quasicrystal state, while the circular nucleation state becomes an index-2 saddle point (Fig.~\ref{fig:s1}D).
After another transcritical bifurcation, the circular nucleation state becomes the transition state (Fig.~\ref{fig:s1}E), where the liquid state becomes the dominant structure.
Note that the lamellar nucleation state will bifurcate successively as $\alpha$ increases, and we did not trace this solution because its index keeps increasing.
We plot the bifurcation diagram to illustrate that for each set of parameters, we can find a transition state connecting the liquid state and the quasicrystal state.
From this transition state, two gradient flows can reach the liquid state and the quasicrystal state.
The results also demonstrate that the critical nuclei may have different sizes and amplitudes with different model parameters.
}

\begin{table}
\centering
\caption{Approximate errors of proper domain sizes for the two-dimensional 12-fold quasicrystals.}
\begin{tabular}{ccccccc}
    \toprule
    approximate error & 0.052559 & 0.019238 & 0.014083 & 0.005155 & 0.003774 & 0.001381\\ \midrule
    $L$  & 22 & 30 & 82 & 112 & 306 & 418\\
    \bottomrule
\end{tabular}\label{tab:sda}
\end{table}

\FloatBarrier

\movie{The transition pathway from the liquid state to the 12-fold quasicrystalline state. The parameters are $\varepsilon=-0.01$, $\alpha=1$, $L=112$, and $N=1024$.}

\movie{The one-stage transition pathway from the 12-fold quasicrystalline state to the 6-fold crystalline state. The parameters are $\varepsilon=0.05$, $\alpha=1$, $L=112$, and $N=1024$.}

\movie{The two-stage transition pathway from the 12-fold quasicrystalline state to the 6-fold crystalline state, mediated by a metastable lamellar quasicrystalline state. The parameters are $\varepsilon=0.05$, $\alpha=1$, $L=112$, and $N=1024$.}

\movie{The one-stage transition pathway from the 6-fold crystalline state to the 12-fold quasicrystalline state. The parameters are $\varepsilon=5\times 10^{-6}$, $\alpha=\sqrt{2}/2$, $L=112$, and $N=1024$.}

\movie{The two-stage transition pathway from the 6-fold crystalline state to the 12-fold quasicrystalline state, mediated by a metastable lamellar quasicrystalline state. The parameters are $\varepsilon=5\times 10^{-6}$, $\alpha=\sqrt{2}/2$, $L=306$, and $N=3072$.}

\bibliography{bib}